\documentstyle[sprocl]{article}

\def\de{\delta}

\def\ka{\kappa}

\def\om{\omega}
\def\Ga{\Gamma}
\def\De{\Delta}

\def\La{\Lambda}

\def\Ps{\Psi}

\def\half{{\textstyle{1\over 2}}}
\def\frac#1#2{{\textstyle{{#1}\over {#2}}}}
\def\lsim{\mathrel{\rlap{\lower4pt\hbox{\hskip1pt$\sim$}}
    \raise1pt\hbox{$<$}}}
\def\gsim{\mathrel{\rlap{\lower4pt\hbox{\hskip1pt$\sim$}}
    \raise1pt\hbox{$>$}}}
\def\sqr#1#2{{\vcenter{\vbox{\hrule height.#2pt
         \hbox{\vrule width.#2pt height#1pt \kern#1pt
         \vrule width.#2pt}
         \hrule height.#2pt}}}}

\def\Re{\hbox{Re}\,}
\def\Im{\hbox{Im}\,}

\def\cL{{\cal L}}

\def\cF{{\cal F}}

\newcommand{\beq}{\begin{equation}}
\newcommand{\eeq}{\end{equation}}
\newcommand{\bea}{\begin{eqnarray}}
\newcommand{\eea}{\end{eqnarray}}

\renewenvironment{thebibliography}[1]
 { \rm
   \begin{list}{\arabic{enumi}.}
    {\usecounter{enumi} \setlength{\parsep}{0pt}
     \setlength{\itemsep}{3pt} \settowidth{\labelwidth}{#1.}
     \sloppy
    }}{\end{list}}

\def\k{$K^0$}
\def\kbar{$\overline{K^0}$}

\begin{document}

\title{CLASSICAL ANALOGUE MODELS FOR T AND CPT VIOLATION}

\author{\'AGNES ROBERTS}

\address{Indiana University Physics Department,\\ Bloomington IN 47405, USA\\E-mail: agrober@indiana.edu} 

\maketitle\abstracts{ Classical-mechanical oscillating systems are analyzed to model
CP violating neutral kaon oscillations.  Through the identification of the key features
of the quantum effective hamiltonian we search for similarities in the characteristic matrix
of the classical systems.  Some no-go results are found for undamped systems
and for systems using two one-dimensional coupled oscillators.  A model
is presented emulating both T and spontaneous CPT violation.}

\section{Introduction}
Among different quantum oscillating systems, the neutral meson systems are
some of the most interesting.  Long-standing experimental results indicate that
the neutral kaons display a small violation of the
combined (C) charge conjugation and (P) parity symmetries.\cite{ccft}
Neutral kaons are now experimentally known to violate (T) time reversal 
symmetry\cite{sachs} as well.
The formalism for the kaon system, however,
could also allow CP breaking via the violation of the combined CPT symmetry\cite{lw}
resulting from spontaneous CPT breaking.\cite{kp2,kp}

One way to analyze these symmetry violations is to model the kaons with a classical
analogue system.  Simple systems of coupled classical harmonic oscillators can be
constructed to parallel the characteristics of the effective hamiltonian describing
the kaon oscillations.  The kaons and antikaons can be modeled, for instance,
using two damped one-dimensional harmonic oscillators
weakly coupled through a damped coupling.\cite{bw}  Even though this represents a system
which demonstrates
the features of the CP eigenstates well, it yields no simple way to model the
CP-violating physical states.

Indeed we'll show in the course of this paper that T violation cannot be modeled via two
one-dimensional oscillators in a natural way.
The only systems able to model kaon oscillations are those with a single two-dimensional
oscillator, an idea also suggested by Rosner.\cite{rs}
In this paper, key features of various classical systems,
with corresponding properties of the quantum system, will
be analyzed, to create a suitable model for T and CPT violation.
It will be shown that models with no damping or models with two coupled
one-dimensional oscillators cannot fully parallel the characteristic
time development of the neutral kaon states.

\section{Basics}

One goal of this section is to discuss convenient formalisms and basic results.  Looking
at these results, we can then create a classical analogue model.
Following the discussion of the kaon system,
the general form of a desired classical model is given.

The general neutral kaon state has a time evolution described by an effective
Hamiltonian $\La$ as $i\de_t\Ps = \La\Ps$.
The physical kaon states $K_S$, $K_L$ are the eigenvectors of $\La$.  While $\La$ consists of
two Hermitian matrices, $M$ the mass matrix and $\Ga$ the decay matrix, the combination
$\La = M - \half i\Ga$ is nonhermitian.  This results in a nondiagonalizable $\La$,
the eigenstates of which overlap and
mix the definite CP eigenstates, leading to CP violation.

This feature can be conveniently formulated using the parametrization\cite{lw}
\begin{equation}
\La = \left(
\begin{array}{lr}
- iD + E_3 & E_1 - iE_2 \\ & \\
E_1 + iE_2 & -iD - E_3
\end{array}
\right),
\label{laparm}
\end{equation}
where $D_1$, $E_1$, $E_2$, $E_3$ are complex.

In terms of the above parameters, the condition for T violation is
\begin{equation}
(\Re E_2\Im E_1 - \Re E_1\Im E_2)\neq 0,
\label{tviol}
\end{equation}
and for CPT violation
\begin{equation}
\Re E_3 \neq 0  \qquad \hbox{or} \qquad \Im E_3 \neq 0.
\label{cptviol}
\end{equation}

This yields three independent real quantities for the determination of CP violation in the
case of the neutral kaons.

Change in the strong phase of \k~ and \kbar~ can mix $E_1$ and $E_2$ without changing
physical behavior.  Condition (2), however, is phase invariant.\cite{kr}

Also, the location of the T and CPT violating parameters can change
by choice of a unitarily equivalent new base to the \k~-\kbar~basis.\cite{rs}
Note, however, that neither of these transformations can mix elements of the mass
matrix $M$ and the decay matrix $\Ga$.

The essential property of the effective hamiltonian is that the CP eigenstates are mixed in the physical
states $K_S$, $K_L$ because of the particular way $M$ and $\Ga$ are combined in it.

In modeling the system classically, first we identify \k-\kbar~with the two generalized coordinates
of a classical harmonic oscillator system.

The individual oscillating frequencies and dampings associated with these two coordinates model the
kaon mass energy and decay rate.  The equal \k-\kbar~masses
and decay rates thus are represented with oscillators, initially having equal
frequency and damping.

To parallel the idea of spontaneously broken CPT symmetry in the context of conventional quantum
field theory\cite{kp2,ck} the classical model should also allow an offset of the originally equal frequencies
and/or damping of the two generalized coordinates arising dynamically with the full motion of the system.
We expect to introduce into our model the analogue for T violation by coupling the classical
oscillators in an appropriate way.  This violation can also be emulated 
spontaneously by the dynamics of the system.

For simplicity, only those models with small oscillations about equilibrium  having linear equations
of motion are discussed here.  Motion of the linear generalized coordinates
of such systems can be described by the two component form $Q=\Re[Ae^{i\om t}]$ with complex
$A$.  In matrix form, the equation of motion is written as $AX = O$.  $X$ is the classical characteristic
matrix.  The intention is to create a classical model, the $X$ matrix of which carries features similar to
those of $\La$.

\section{Models without damping}

In this section discussing undamped systems, we will establish several no-go results.

The Lagrangian of a conservative classical-mechanical system of small oscillations about equilibrium can be
expanded into a quadratic form.  The systems considered here have two degrees of freedom, with real
generalized coordinate $Q(t)$.  A general Langrangian
of this kind has the form\cite{lm}
\begin{equation}
\cL = \half \dot Q ^T T \dot Q
+ \half \dot Q^T G Q
- \half Q^T V Q,
\end{equation}
where $T$, $G$, $V$ are 2x2 matrices.
$T$,$V$ are the symmetric kinetic and potential energy matrices while G, the gyroscopic matrix, is always antisymmetric.\cite{lm}
The equation of motion of the system is $T\ddot{Q} + G\dot{Q} + VQ = 0$.
Substituting the solution for Q, the characteristic
equation is $XA = 0$, giving the characteristic matrix as
\begin{equation}
X = -\om^2T + i\om G + V
\end{equation}

Inspecting $X$, we see that this type of linear nondissipative system has hermitian
characteristic matrices, the eigenvectors of which cannot mix.  Hence $X$ cannot model
the nonhermitian effective hamiltonian of the kaons.
A term-by-term comparison between the
elements of $X$ and those of $\La$ also shows that all eight parameters
of $\La$ cannot be paralleled with a nondissipative system.\cite{kr}
Similarly, T violation in the neutral kaon system involves dissipative oscillations.

Comparing $X$ and $\La$, we find that $\Re E_2$ is emulated by the gyroscopic term.
This property carries
over into dissipative systems as well.
This motivates us to consider only systems with a gyroscopic matrix.
We find, however, that a system with two degrees
of freedom with two one-dimensional coupled oscillators cannot produce such terms.\cite{kr}
To generate the gyroscopic term, we consider models with time-dependent (rheonomic) constraints.  In particular, we look
at a model with a uniformly rotating constraint.

\section{Model with viscous damping}

Based on the considerations of the previous sections, we now present a model for both T and CPT violation.
Take as our model the surface of a bowl spherical on the bottom.  A particle oscillates under gravity
on the bottom with equal frequencies in two orthogonal directions.  With increased heights the cross section of this bowl
smoothly transforms into a uniform elliptical shape.  When the bowl is rotated with a small amount of friction present,
the particle climbs to some equilibrium height determined by gravity and the frequency of rotation.  This equilibrium
for the elliptical cross section is at either side of the major axis.

After assuming its equilibrium position, the particle rotates with the bowl.  The small oscillations it executes vertically
and horizontally are both stable.  These two oscillations, coupled by the rotation of the bowl, generate a Lagrangian of the
form of equation (4), including the gyroscopic matrix.

From the discussion above, we
know that dissipation needs to be introduced into the model.  To keep the equations of motion linear, we only consider
damping describable by the general quadratic expression\cite{lm}
\begin{equation}
\cF = \half \dot Q^T R \dot Q + \dot Q ^T H Q.
\end{equation}
Here, the first term is the Rayleigh dissipation function.  The second refers to linear homogeneous damping forces proportional to
generalized coordinates.\cite{lm}  The modified equation of motion including damping has the form
\begin{equation}
T \ddot Q + (G+R) \dot Q + (V+H) Q=0 .
\end{equation}

For the example of the bowl, both such damping terms can be introduced as external viscous damping forces.
We can imagine, for instance, a mesh bowl rotating in a static fluid.  The fluid exerts resistive force on the particle oscillating
inside.  This system will produce the desired characteristic matrix\cite{kr}
\begin{equation}
X = -T\om^2 + i(G+R)\om + V + H.
\label{x}
\end{equation}

Here $T,V,R$ are real symmetric and $G,H$ real antisymmetric matrices.\cite{lm}  The $\om$ of the damped system is complex.  Taking
$\om=\mu+i\ka$ and defining $\om^2=\De^2+2i\mu\ka$, where $\De^2=\mu^2-\ka^2$, a substitution of $\om$ into equation (8) gives
$X$ to be the sum of the hermitian matrices $-T\De^2 +i\mu G - \ka R + V$
and the antihermitian matrices $-2i\mu\ka T + i\mu R - \ka G + H$, which is overall nonhermitian.
Hence $X$ is nondiagonalizable and has coupled eigenstates.  Detailed inspection reveals that there is a difference between diagonal
elements of $X$, generating a correspondence to both $\Re E_3$ and $\Im E_3$ of $\La$.  This difference is generated by the rotation of
the bowl, emulating spontaneous CPT violation.  $X$ is found to produce an analogue for all eight parameters of $\La$.\cite{kr}

\vskip-1pt
\section{Summary}
\vskip-1pt

This paper studied classical oscillating systems to parallel CP violating neutral kaon oscillations.  We found that analogue
model without damping and ones with two one-dimensional oscillators cannot be sufficient models.  We gave an example showing that
properly damped two-dimensional oscillators with time-dependent constraint can indeed model spontaneous CPT and T violation.

\end{document}